\documentclass[sigconf]{acmart}

\copyrightyear{2024}
\acmYear{2024}
\setcopyright{acmlicensed}\acmConference[CHI '24]{Proceedings of the CHI Conference on Human Factors in Computing Systems}{May 11--16, 2024}{Honolulu, HI, USA}
\acmBooktitle{Proceedings of the CHI Conference on Human Factors in Computing Systems (CHI '24), May 11--16, 2024, Honolulu, HI, USA}
\acmDOI{10.1145/3613904.3642462}
\acmISBN{979-8-4007-0330-0/24/05}

\usepackage{exii-macros}

\newcommand{\name}{DirectGPT\xspace}

\newcount\S \newcount\H  \newcount\Hrest  \newcount\M  \newcount\Mrest

\newcommand{\prettytimestamp}[1]{%
   \setS#1.\end
   \H=\S \divide\H by3600
   \Hrest=\H \multiply\Hrest by-3600 \advance\Hrest by\S
   \M=\Hrest \divide\M by60
   \Mrest=\M \multiply\Mrest by-60 \advance\Mrest by\Hrest
   \ifnum \H=0
     \ifnum\M=0 \miliseconds#1s...\end             
     \else \the\M min \the\Mrest s\fi            
   \else \the\H h\the\M min\the\Mrest s\fi 
}
\def\setS#1.#2\end{\S=#1\relax}
\def\miliseconds#1.#2#3#4\end{#1\ifx.#2\else.#2\ifx.#3\else#3\fi\fi}

\ExplSyntaxOn
\NewExpandableDocumentCommand{\round}{m}
{
	\fp_eval:n { round(#1,0) }
}
\ExplSyntaxOff

\usepackage{stfloats}

\usepackage{cleveref}

\usepackage{setspace}

\newenvironment{dialogue}{
\par
  \noindent
  \ttfamily
  \setlength{\parindent}{0pt} 
  \setlength{\leftskip}{20pt}  

}{
  \par
  \medskip
}

\begin{document}


\title{DirectGPT:  A Direct Manipulation Interface to Interact with Large Language Models}

\author{Damien Masson}
\orcid{0000-0002-9482-8639}
\affiliation{%
  \institution{Cheriton School of Computer Science, University of Waterloo}
  \city{Waterloo}
  \country{Canada}
}
\additionalaffiliation{\institution{University of Toronto}
  \city{Toronto}
  \country{Canada}
}
\email{dmasson@uwaterloo.ca}

\author{Sylvain Malacria}
\orcid{0000-0002-5201-5875}
\affiliation{
  \institution{Univ. Lille, Inria, CNRS, Centrale Lille, UMR 9189 CRIStAL}
  \postcode{F-59800}
  \city{Lille}
  \country{France}
  }
\additionalaffiliation{%
  \institution{University of Waterloo}
  \country{Canada}}
\email{sylvain.malacria@inria.fr}

\author{G\'ery Casiez}
\orcid{0000-0003-1905-815X}
\affiliation{%
 \institution{Univ. Lille, CNRS, Inria, Centrale Lille, UMR 9189 CRIStAL}
 \postcode{F-59000}
 \city{Lille}
 \country{France}}
\additionalaffiliation{%
 \institution{Institut Universitaire de France}
 \city{Paris}
 \country{France}}
  \additionalaffiliation{%
  \institution{University of Waterloo}
  \country{Canada}}
\email{gery.casiez@univ-lille.fr}

\author{Daniel Vogel}
\orcid{0000-0001-7620-0541}
\affiliation{%
  \institution{Cheriton School of Computer Science, University of Waterloo}
  \city{Waterloo}
  \country{Canada}
}
\email{dvogel@uwaterloo.ca}



\begin{abstract}
We characterize and demonstrate how the principles of direct manipulation can improve interaction with large language models. This includes: continuous representation of generated objects of interest; reuse of prompt syntax in a toolbar of commands; manipulable outputs to compose or control the effect of prompts; and undo mechanisms. This idea is exemplified in DirectGPT, a user interface layer on top of ChatGPT that works by transforming direct manipulation actions to engineered prompts. A study shows participants were 50\% faster and relied on 50\% fewer and 72\% shorter prompts to edit text, code, and vector images compared to baseline ChatGPT. Our work contributes a validated approach to integrate LLMs into traditional software using direct manipulation. Data, code, and demo available at \url{https://osf.io/3wt6s}.

\end{abstract}

%
%

\ccsdesc[500]{Human-centered computing~Interactive systems and tools}
\ccsdesc[500]{Human-centered computing~Interaction design theory, concepts and paradigms}

\keywords{direct manipulation, large language models, prompt engineering}

 \begin{teaserfigure}
   \includegraphics[width=\textwidth]{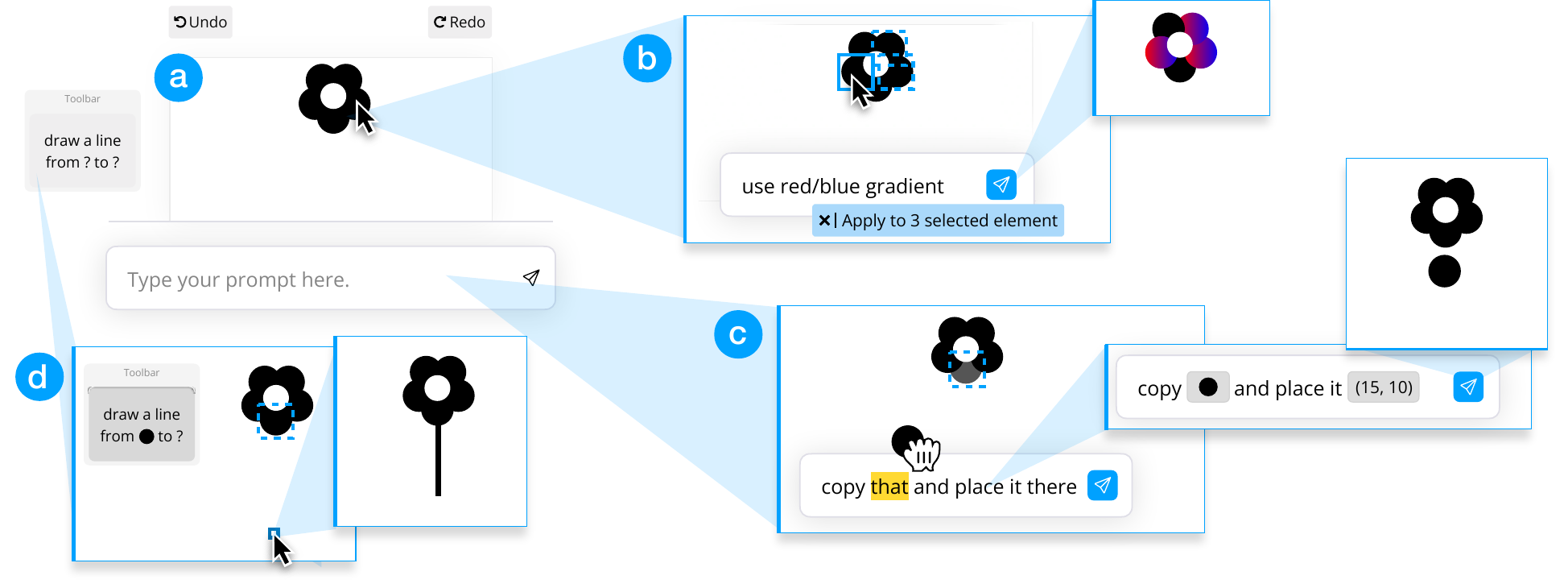}
   \caption{DirectGPT used on a vector image to demonstrate direct manipulation principles for LLMs: (a) continuous representation of the objects of interest; physical actions to (b) localize the effect of prompts and (c) refer to objects; (d) reusable prompts in a toolbar of commands; and reversible operations through undo and redo features.}
   \Description{Screenshots of the DirectGPT interface with a drawing of a petal, undo and redo buttons, a prompt field, and a toolbar. Starting from the drawing of a flower, the user can either select some of the petals and apply a prompt; drag a petal in the prompt field; or click on the tool from the toolbar to reuse a previous prompt as an ad hoc tool.}
   \label{fig:teaser}
 \end{teaserfigure}

\maketitle



\section{Introduction}
Given a textual instruction in the form of a ``prompt'', a large language model (LLM) can generate outputs such as emails, computer code, and vector images.
The output often needs to be tweaked by conversing with the LLM until obtaining a satisfactory result. However, this tweaking process uses relatively slow text input, it requires precise references to elements in the output, and the wording must be carefully chosen to workaround unintuitive limitations of LLMs~\cite{reynoldsPromptProgrammingLarge2021, openaiGPTBestPractices2023, zamfirescu-pereiraWhyJohnnyCan2023}. 

These issues are reminiscent of the reasons why \textit{direct manipulation interfaces} emerged as an alternative to \textit{command line interfaces}~\cite{shneidermanDirectManipulationStep1983, hutchinsDirectManipulationInterfaces1985, shneidermanFutureInteractiveSystems1982}.
The problems identified 40 years ago also apply to interfaces for LLMs today:
(i) \textit{indirect engagement} due to manipulating natural language instead of the objects of interest; (ii) \textit{semantic distance} due to the verbosity and difficulty to convey intent in a prompt; and (iii) \textit{articulatory distance} due to the form of prompts being poor representations of the actions they are meant to convey.
In fact, interfaces for LLMs like OpenAI's ChatGPT~\cite{openaiIntroducingChatGPT2022}, Google's Bard~\cite{googleBardChatBased2023}, and Microsoft's Bing~\cite{microsoftIntroducingNewBing2023} undermine most of what makes an interface direct.
They rely on a linear textual history instead of continuous representation of the objects of interest; carefully crafted prompts remain in the history instead of commands that are rapid, incremental, and reversible;
and complex language is manipulated instead of the objects of interest.
As a result, current interaction with LLMs lacks most of the benefits of direct manipulation such as improved learnability, speed of execution, goal feedback, and error prevention and recovery~\cite{shneidermanDirectManipulationComprehensible1997, shneidermanFutureInteractiveSystems1982, hutchinsDirectManipulationInterfaces1985}.

Consider the task of tweaking a generated SVG drawing.
Directly pointing at shapes and locations using a mouse cursor is much easier than specifying them using words that may rely on metaphors or geometric references. Besides being verbose, these high-level descriptions are quite far from the SVG code that the model generates.
There is also ambiguity if the shape appears multiple times or if the wording assumes spatial information that does not align with how the model represents the drawing.
Moreover, applying the same transformation to a different shape requires going through this process again.
These difficulties apply when tweaking generated text as well. Consider how replacing a specific word with a synonym in a body of text requires a careful description of the word to change and its position, the action to be done, and other constraints such as keeping the rest of the text intact.
Regardless of what is being generated, mistakes are costly as incorrect results are hard to notice and they remain in the conversation history, possibly impacting future interactions.

Instead of solely relying on words, we introduce and characterize prompting through direct manipulation as a way to facilitate conversations with LLMs.
Directness is created by interacting with the output to refer to specific objects when crafting new prompts (\cref{fig:teaser}c), or to localize the desired area of effect for the prompt (\cref{fig:teaser}b).
Reversibility is realized through typical undo and redo features instead of long linear conversations.
Immediate feedback is provided by highlighting localized changes and elements about to be modified.
And rapid reusable commands are generated by abstracting previous prompts into templated snippets (\cref{fig:teaser}d).

\rev{To explore the benefits of these principles for LLMs, we implemented them into a prototype system called DirectGPT.}
In a user study we compare DirectGPT to ChatGPT for text, code, and vector image editing tasks, and found that participants relied and preferred the mechanisms of direct manipulation when accomplishing specific and localized tasks.
Specifically, participants used 50\% fewer and 72\% shorter prompts all while being 50\% faster and 25\% more successful at accomplishing tasks.
\rev{Beyond our system and its interactions, our findings inform the design and integration of LLMs and other prompt-driven AI into graphical user interface applications.}

\section{Background and Related Work}
Direct manipulation is often opposed to intelligent agents when debating controllable systems~\cite{shneidermanIntelligentMachinesJust1993} versus autonomous systems~\cite{maesAgentsThatReduce1994}.
This work takes the perspective that both can work together to enable \textit{``increased automation that amplifies the productivity of users and gives them increased capabilities in carrying out their tasks, while preserving their sense of control and their responsibility''}~\cite{heerAgencyAutomationDesigning2019, shneidermanDirectManipulationVs1997}.
We begin by defining what we mean by \textit{direct manipulation} and then review issues with LLMs and how prior work has attempted to solve them through more direct interfaces.

\subsection{What is Direct Manipulation?}
Direct manipulation is a term coined by Ben Shneiderman to describe characteristics of interactive systems that were found to \textit{``generate glowing enthusiasm among users''}~\cite{shneidermanDirectManipulationStep1983, shneidermanFutureInteractiveSystems1982}. 
Shneiderman found these systems to share three principles:
(1) \textit{``Continuous representation of the object of interest''}, ideally in its final form~\cite{shneidermanDirectManipulationStep1983}; (2) \textit{``Physical actions or labelled button presses instead of complex syntax''}, especially if they involve visual selection of objects of interest~\cite{hutchinsDirectManipulationInterfaces1985}; and (3) \textit{``Rapid, incremental, reversible operations whose impact on the object of interest is immediately visible''}~\cite{shneidermanDirectManipulationStep1983}.
We use the term \textit{direct manipulation} to refer to these principles, and we consider interfaces to be more or less \textit{direct} depending on how thoroughly they implement these principles.

\subsection{Five Issues With Prompting that Motivate the use of Direct Manipulation}
First, \textit{intent is hard to convey with prompts} due to language being ambiguous in many tasks.
This is particularly true for structured information such as graphs and tables~\cite{liuPretrainPromptPredict2021} and information poorly described in words, such as images~\cite{dangHowPromptOpportunities2022}.
Users need to find the right phrasing that describes the task and that the system understands~\cite{liuWhatItWants2023}.
But, similar to how vocabulary mismatch is the leading cause of failure when new users try to verbalize their goals~\cite{novickMicrostructureUseHelp2009}, users are unlikely to know the best vocabulary to describe certain objects or tasks.
Instead, direct manipulation interfaces reduce ambiguity by referring to objects directly instead of describing them.

Second, \textit{conversation is a poor metaphor to design effective and concise prompts}.
The nature of LLMs leads to a conversational interface and style of AI voice that encourages users to write prompts the way they would address a human.
As a result, some effective prompt engineering strategies such as giving examples or repeating instructions multiple times are rarely used because they are unnatural in human-human conversation~\cite{zamfirescu-pereiraWhyJohnnyCan2023, ohLeadYouHelp2018}.
Additionally, users tend to use polite language such as ``thank you'' or ``can you [...] please?'' which adds to the verbosity of prompts and slows down the interaction~\cite{zamfirescu-pereiraWhyJohnnyCan2023}.
Instead, direct interfaces remove this expectation, and specific prompt engineering strategies can be hidden into the internal functioning of the interface. 

Third, \textit{the effect of a prompt is hard to control and predict} due to producing different and possibly unexpected results after a slight rewording~\cite{luFantasticallyOrderedPrompts2022, vaithilingamExpectationVsExperience2022}.
An example are ``do not'' statements that LLMs are known to ignore, or worse, treat as ``do'' statements~\cite{zamfirescu-pereiraWhyJohnnyCan2023}.
As a result, trying to edit a specific part of the generated output often leads the LLM to modify other parts that should remain untouched.
In contrast, our direct interface helps users convey constraint and enforces them instead of solely relying on the LLM.

Fourth, \textit{prompt engineering requires a lot of trial-and-error} before obtaining a satisfactory result~\cite{dangHowPromptOpportunities2022, dangChoiceControlHow2023}. In contrast to conversational interfaces, direct interfaces facilitate trial-and-error behaviours~\cite{massonSuperchargingTrialandErrorLearning2022, draperLearningExplorationAffordance1993, demulLearningUserInterfaces1996}. For example, manipulating objects and clicking buttons make for fast iterations easily recovered from with undo features.

Fifth, \textit{prompts are hard to reuse} since they are each designed for a single, specific interaction~\cite{dangHowPromptOpportunities2022}.
In a conversational interface, the prompt specifies both the action to be done and the object to modify.
As such, reusing a prompt requires editing part of the describing the object.
In contrast, direct manipulation interfaces typically rely on two distinct steps to specify the action and the object.
Our work describes how this specificity can be leveraged to populate a toolbar of commands that help reuse prompts on new objects.

\subsection{Systems to Help Craft Better Prompts}
When the LLM responds unexpectedly, rewording the prompt may help. To assist in this task, many strategies have been suggested~\cite{openaiGPTBestPractices2023, liuPretrainPromptPredict2021} and prior work proposed systems to help with \textit{prompt engineering}. For example, PromptIDE~\cite{strobeltInteractiveVisualPrompt2022} and PromptAid~\cite{mishraPromptAidPromptExploration2023} provide interactive interfaces to test different prompts, evaluate their performance, and suggest variations. When a single prompt is not enough, AI Chains~\cite{wuAIChainsTransparent2022},  PromptChainer~\cite{wuPromptChainerChainingLarge2022}, and ChainForge~\cite{arawjo2023chainforge} help construct chains of prompts and evaluate their effectiveness.

These systems sometimes offer interfaces to make it easier to adjust prompts, change configurations, and reuse previous prompts and outputs.
However, they are typically used to evaluate and generate prompts to be integrated as part of bigger pipelines to solve recurrent problems.
In contrast, our work uses LLMs to manipulate objects of interest directly such as editing text and images.

\subsection{Prompting through More Direct Interactions}
While interfaces like ChatGPT are strictly conversational, alternative interfaces have included some ideas from direct manipulation to control generative AIs for certain tasks.

\subsubsection{Labelled Buttons Instead of Verbose Prompts}
Most integration of generative AIs rely on widgets such as buttons and sliders to execute a pre-defined prompt or modify parameters of the model.
For example, using sliders to manipulate continuous parameters that impact the generation of images~\cite{dangGANSliderHowUsers2022, rossEvaluatingInterpretabilityGenerative2021} and music~\cite{louieNoviceAIMusicCoCreation2020}.
Buttons have been used as shortcuts to execute predefined prompts that ask for more information about a topic~\cite{jiangGraphologueExploringLarge2023a, suhSensecapeEnablingMultilevel2023a}, edit a story~\cite{yuanWordcraftStoryWriting2022, clarkCreativeWritingMachine2018}, and run macros to generate computer code~\cite{jiangGenLineGenFormTwo2021}. 
Our work also relies on buttons to populate a toolbar of commands, but these are dynamically generated as users interact with the LLM.

\subsubsection{Physical Actions Instead of Verbose Prompts}
Gestures and physical metaphors have been employed to help tweak the output of generative AIs.
For example, pointing at a specific location to refer to elements of an image~\cite{liuInternGPTSolvingVisionCentric2023} or force a model to  re-generate only a specific part of an image~\cite{yuGenerativeImageInpainting2018}.
Similarly, dragging gestures have been used to manipulate spatial attributes of an image such as the pose, facial expression, and layout~\cite{wangImprovingGANEquilibrium2022, panDragYourGAN2023}. Metaphorical gestures in the form of a brushing action were used to control the fortune of  characters in a story~\cite{chungTaleBrushSketchingStories2022a} and express image modifications by mixing prompts through physical actions similar to mixing colours~\cite{chungPromptPaintSteeringTexttoImage2023a}.
Our work highlights how physical actions and tool metaphors can be leveraged for more general prompt-based interaction suitable for diverse application domains.

\subsubsection{Reversible Operations Instead of Continuous Conversations}
Conceptual approaches to interaction  histories have been explored in detail~\cite{nancelCausalityConceptualModel2014}, yet popular interfaces for LLMs are limited in this regard.
For example, with ChatGPT, there is no explicit action to recover from a mistake. Instead, users have the choice to start a new conservation from scratch, edit the conversation history, or compose a prompt asking the model to revert back to a previous state.
For the latter, the revert prompt itself is saved in the history which can impact future interactions~\cite{luFantasticallyOrderedPrompts2022, vaithilingamExpectationVsExperience2022}.
Similar to how traditional software may use subjunctive interfaces~\cite{lunzerSubjunctiveInterfacesExtending2008} and parallel paths~\cite{terryVariationElementAction2004} to help explore and recover from mistakes, recent work proposed node-based and hierarchy-based explorations of the responses generated by LLMs~\cite{suhSensecapeEnablingMultilevel2023a, jiangGraphologueExploringLarge2023a, angertSpellburstNodebasedInterface2023b}. With these interfaces, users can go back to a previous node and start from this point on to further explore. Instead of using current or novel representations of history, our work proposes using traditional undo mechanisms to support reversibility at the granularity of an operation.

\subsubsection{Immediate Feedback Instead of Word-by-Word Generation}
Large generative models often require considerable computing time and power, so generating a result immediately remains an open challenge~\cite{chenNextStepsHumanCentered2023, leviathanFastInferenceTransformers2023}.
Most previous work focuses on providing feedback at the speed of the generation.
For example, ChatGPT streams a response word-by-word as it is generated.
However, such feedback makes it difficult to notice differences between iterations.
To make output feel closer to realtime, LLMs can be prompted to generate intermediary parameterized representations that are fast to recalculate.
For example, instead of generating an image, the LLM can generate code that generates the image. Then, the parameters of the code can be manipulated with immediate feedback~\cite{angertSpellburstNodebasedInterface2023b}.
However, this approach is limited to outputs easily generated through code.
Instead, our work exploits spatial input from direct manipulation to provide immediate loading feedback on elements about to be modified to keep users aware of modifications.

\subsection{Blending Language and Direct Manipulation}
Many systems have sought to combine natural language and direct manipulation to be easier to use.
One of the earliest examples is ``Put-That-There''~\cite{boltPutthatthereVoiceGesture1980} where vocal commands specify actions and pronouns and mid-air pointing gestures specify objects and locations. 
Of course, due to limitations of 1980s technology, the system supported only a small set of commands and sentence structures. 
But conceptually, the approach allows manipulating images without having to learn new software. In fact, Wizard of Oz experiments later showed participants preferred that modality~\cite{hauptmannSpeechGesturesGraphic1989}.

More recently, there has been substantial progress in robust ways to process natural language. 
For example, PixelTone~\cite{laputPixelToneMultimodalInterface2013} relies on algorithms to support a broader set of sentence structures. 
While users were still constrained to the phrase templates and vocabulary understood by the system, an experiment showed the speech and gesture interface was preferred to accomplish photo editing tasks. 
To handle more complex natural language queries, solutions have been proposed to map natural language to user interface commands, such as constructing a graph~\cite{fourneyQueryfeatureGraphsBridging2011} and building vector representations~\cite{adarCommandSpaceModelingRelationships2014}. Alternatively, for the task of programming agents, PUMICE~\cite{liPUMICEMultiModalAgent2019} proposes an iterative refinement of the initial natural language query with possibilities to demonstrate actions by direct manipulation whenever a request is misunderstood.

Arguably the closest approach to our work is Stylette~\cite{kimStyletteStylingWeb2022}, a system to customize websites using speech and selection of elements.
It uses a pipeline of deep learning models to support better understanding of natural language queries and suggest CSS properties to modify.
\rev{Similarly, our prototype system is built on top of an LLM. However, our work is not targeted at a specific task, instead it explores how the principles of direct manipulation can be extended to LLMs to support tasks such as text, code, and image editing.}
The design we propose is specifically adapted to address limitations of LLMs, such as easing verification, better control, and facilitating trial-and-error. We also cover all aspects of direct manipulation, not just selection mechanisms, and study their benefits compared to relying solely on natural language. In doing so, we introduce mechanisms, such as prompt reuse by turning previous prompts into templated snippets that populate a toolbar of commands.

\section{\name: an exemplar direct interface for LLMs}\label{sec:directgpt}
\begin{figure*}[t]
    \includegraphics[width=\textwidth]{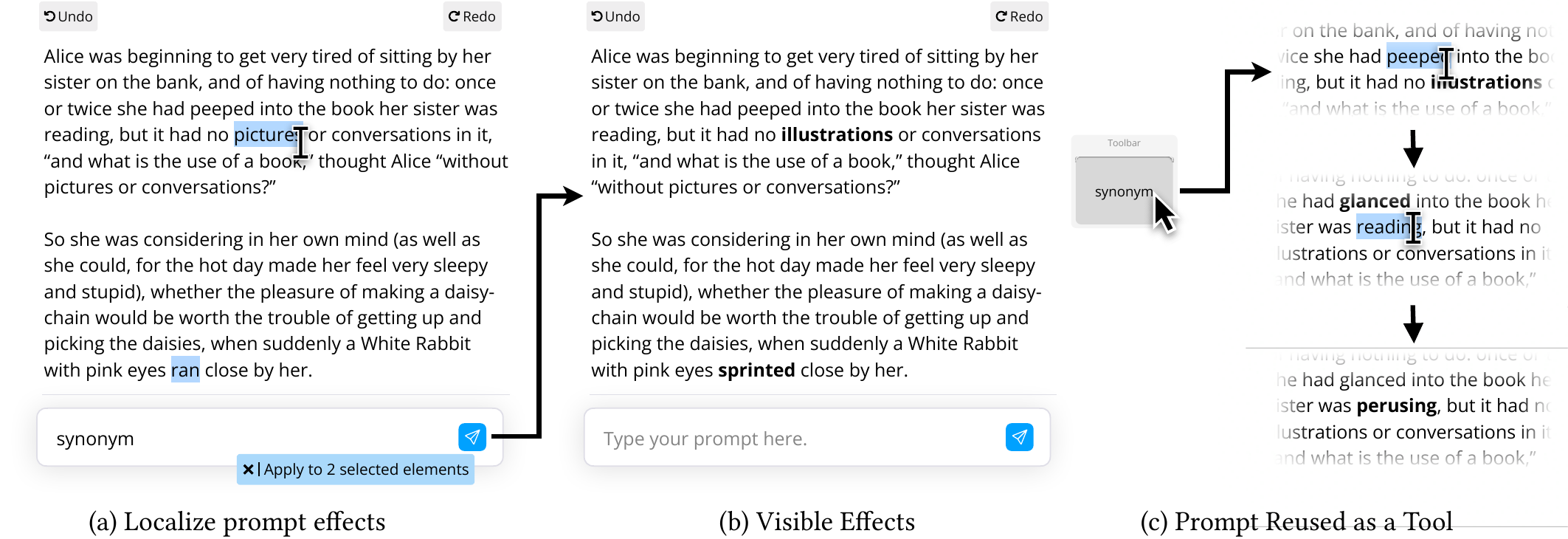}
    \caption{\name used to replace words with synonyms: (a) selecting an object such as a word before writing a prompt forces the prompt to apply to only this object, localizing its effect; (b) once executed, the modifications are highlighted and the ``synonym'' ad hoc tool is created; (c) the ``synonym'' tool is used to quickly find synonyms for other words.}
    \label{fig:synonym}
    \Description{Screenshots of a scenario in which DirectGPT is used to find synonyms. First the two words ``pictures'' and ``ran'' are selected and the prompt ``synonym'' applied. Then the tool ``synonym'' is created and used to find synonyms for more words.}
\end{figure*}

This section describes \name, a direct manipulation interface to interact with LLMs.
To demonstrate the utility of \name, we first illustrate its functionalities in a use case scenario. 
We then follow it with a detailed description of the proposed direct manipulation mechanisms and their implementation.

\subsection{Example Use Case}
Sam is working on a story and would like to revise it before sharing it with friends. 
Sam starts \name and pastes the current version of the story. 
Sam decides that a character, the ``White Rabbit'', needs a vivid description when introduced. Sam selects ``a White Rabbit'' with the mouse cursor, types ``add description of its tail'' in the prompt field, and presses enter.
The selected text pulses and then is replaced by ``a White Rabbit with a fluffy, cotton-like tail''. 
Next, Sam notices that the words ``pictures'' and ``ran'' are used repetitively. Sam selects the first instance of both words (by pressing ctrl while selecting the two words with the cursor) and then types ``synonym'' (\cref{fig:synonym}a).
Again, the two words pulse before being replaced by ``illustrations'' and ``sprinted'' (\cref{fig:synonym}b).
Later, Sam realizes other words should be substituted.
Sam clicks on the ``synonym'' button in the toolbar that appeared after running the previous prompt.
With the cursor now indicating ``synonym'', Sam selects a word and it pulses and then is replaced by a synonym (\cref{fig:synonym}c). Sam continues by selecting other words to edit using this ad hoc ``synonym tool''.

\begin{figure*}[t]
    \includegraphics[width=\textwidth]{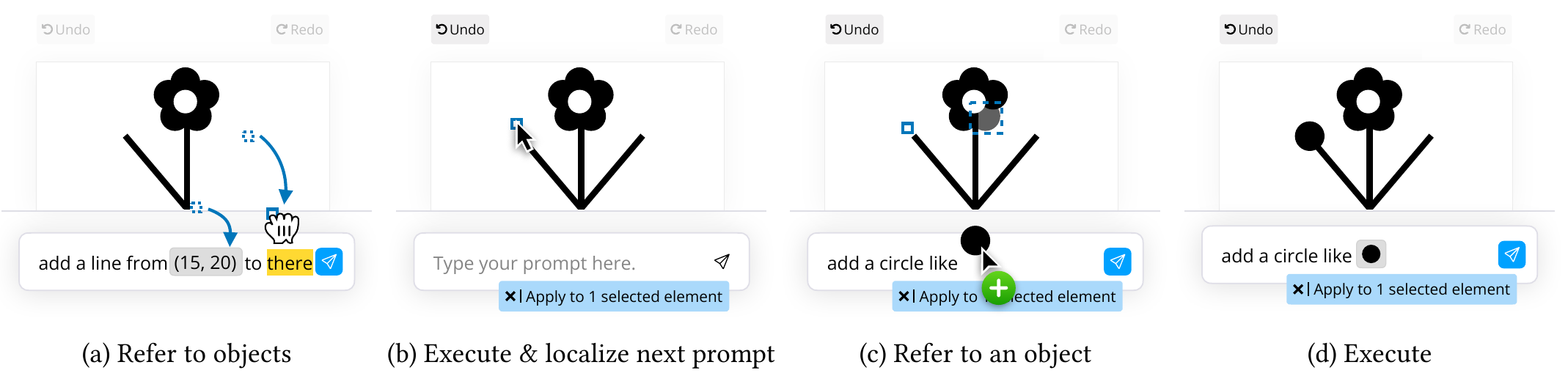}
    \caption{\name used to finish drawing a flower: (a) draw a line by referring to specific locations through drag-and-drop; (b) the line is drawn, click to specify where to add the circle; (c) refer to another circle to copy its size; (d) the circle is added.}
    \label{fig:flower}
    \Description{Screenshots of a scenario in which a flower is being drawn. First, one of the leaf is drawn by prompting to draw a line and dropping locations in the prompt. Then one of the petal is dropped in the prompt to copy it and place it at the top of the leaf.}
\end{figure*}

Later, Sam decides to illustrate the story with a picture.
In DirectGPT, Sam types ``draw a flower with 6 petals''. An image with six circles is generated (five black petals in a circle, and a white circle in the middle).
This is a good start, but it lacks a stem and leaves.
Sam types ``draw a black line from here to there''.
Then, Sam clicks the bottom petal, drags it, and drop it on the word ``here'' in the prompt. Immediately, the word ``here'' is replaced by a picture of the petal. 
Next, Sam drags a pixel located near the bottom of the image and drops it on ``there'' which  is replaced by the corresponding pixel coordinates (\cref{fig:flower}a).
After pressing enter, the bottom pedal and pixel location start pulsing until a line is drawn.
Next, Sam presses the ``draw a black line from ? to ?'' button that just appeared in the toolbar.
With this ad hoc tool active, Sam clicks on two pixel locations, and a line is drawn. They repeat the operation to create a symmetric branch on the other side of the flower.
Finally, Sam clicks the top of the first line to localize the edit, then types ``add a circle like this'', and drags and drops one of the petals onto``this'' (\cref{fig:flower}c).
The operation is repeated on the other side to obtain the final result (\cref{fig:flower}d).

\subsection{Direct Manipulation Principles for LLMs}
The direct manipulation of an element relies on commands (e.g., buttons in ribbons and toolbars) applied to objects of interest (e.g., words of a document in a word processor, elements on a canvas in a drawing software).
This type of operation is commonly known as a \textit{verb} and \textit{noun} construction~\cite[ch. 3-3]{raskinHumaneInterfaceNew2000}.
While the order and interactions to specify the verb and noun might differ, it is typical that clicking a command specifies the verb and pointing at the objects specifies the noun~\cite{wolfTaxonomicApproachUnderstanding1987}.
A parallel can be drawn with conversational interfaces like ChatGPT: the object of interest is the last generation from the model and the commands are generated with the prompt field. However, there is no equivalent to the noun-verb construction. Instead, both noun and verb are specified in a singular prompt by interacting solely with the prompt field.

Below, we review the principles of direct manipulation and show how to leverage them for LLMs. Each principle is demonstrated using DirectGPT, an exemplar direct interface for LLMs. DirectGPT is designed to preserve the standard prompting capabilities of ChatGPT while introducing a layer of complementary interaction mechanisms that are expressive and consistent across application domains such as editing text, code, or images.

\subsubsection{Continuous Representation of the Last Output}
At its core, direct manipulation is possible only if the objects of interest are continuously represented and accessible at all times~\cite{beaudouin-lafonInstrumentalInteractionInteraction2000}. 
This might be the biggest difference with conversational interfaces because it requires moving away from long textual explanations to instead display an object that remains at the same position.
As a result, the interface does less telling and more showing: instead of the model explaining the changes, they should be noticeable through spatial feedback and temporal changes, making it possible to compare the previous state with the new state~\cite{hutchinsDirectManipulationInterfaces1985}.

Another crucial point is that objects should be represented in a final form matching the user's intent~\cite{shneidermanDirectManipulationStep1983}. For example, if users wish to wrangle data, then the representation should show transformed data instead of code to transform it. 
Similarly, if users wish to modify a vector image, then the system should render this image.

DirectGPT shows the object of interest such as code, text, and rendered images at a fixed position on the screen (\cref{fig:flower}). Every new operation generating a response from the model updates this representation to help users notice the differences (\cref{fig:synonym}b and \cref{fig:flower}b).

\subsubsection{Physical Actions Through Prompt-Object Interactions}\label{sec:physical_actions}
Direct interfaces use physical actions and simple metaphors to convey intent without words~\cite{shneidermanDesigningUserInterface2017}.
In our case, the objects of interest and the prompt field provides two interaction contexts with possibilies for physical actions that combine these contexts.
In DirectGPT, prompts can be entered by typing in the prompt field to preserve the capabilities of prompt-based interfaces like ChatGPT.
Additionally, we describe below how interactions with objects can formulate and constrain prompts through physical actions.
Essentially, these interactions recreate the verb-noun constructions typical of direct manipulation interfaces.

\medskip\noindent\textit{Prompt to Object: Localizing Effects --}
Interactions between the prompt field and objects enables specifying objects without having to refer to them in the prompt.
This helps users express intents that involve an object and apply only to that object.
As such, the prompt needs only to describe the verb whereas the physical action unambiguously specifies the noun.
This interaction also describes a constraint in that only the specified objects should be modified while the rest of the content should remain unchanged.
For example, colouring object A requires selecting A and prompting ``red'' or ``colour red''.

In DirectGPT, objects are selected with a click.
When one or multiple objects are selected, the  prompt field indicates ``Apply to selected elements'' (\cref{fig:synonym}a).
Executing a prompt while objects are selected forces the prompt to apply only to these objects. Otherwise, the prompt applies to the whole content, like ChatGPT. This reuse of the prompt field allows users to seamlessly switch between local and global prompts, even after having started typing a prompt.
To maintain compatibility with different domains, the interaction supports both noun-verb (frequent in text editing tasks) and verb-noun constructions (frequent in image editing tasks). This  corresponds to the difference between selecting and then typing the prompt, or typing a prompt and then selecting.

\medskip\noindent\textit{Object to Prompt: Referring to Objects --}
Interactions between objects and the prompt field allow unambiguous references to objects.
This helps users express intents that involve relationships between objects and that may modify the whole content. 
Unlike regular prompt-based interfaces, target objects do not need to be described with words, but are specified instead through a physical action.
This is similar to previous work showing an increase in expressivity and precision when using mid-air pointing gestures to refer to graphical shapes~\cite{boltPutthatthereVoiceGesture1980}, and incorporating images into code to refer to UI components~\cite{yehSikuliUsingGUI2009}.

In DirectGPT, we adopt a drag-and-drop approach to select one or multiple objects and drop them in the prompt. 
Objects can be dropped between words (\cref{fig:flower}c), or onto words (\cref{fig:flower}a) to clarify intention.
This supports both back-and-forth interaction (e.g., type ``move'', drop an object to form next word, type ``to'', drop a location to form another word) or prompt-then-bind interaction (e.g., type ``put that there'', then drop an object onto``that'', and a location into ``there'').
Both interactions have the same effect but one gives users the option to type the whole prompt without being interrupted.
The objects dropped in the prompt are then treated as ``object-words'': their background becomes grey and they can be copied or deleted like single words.
Furthermore, to reduce articulatory distance, DirectGPT uses a thumbnail of the object, or a brief description to render the object-word within the prompt (\cref{fig:flower}a). Hovering over an object-word highlights the corresponding object in the generated output.

\medskip\noindent\textit{Object to Object: Fine-grained Manipulations --}
Interactions on objects support fine-grained manipulation.
This helps users express transformations that involve an object and apply only to this object.
Typically, these interactions are used for common domain-specific edits, such as resizing and moving a shape, indenting a piece of code, or removing some text.
It would be possible to implement such interactions using a task-specific approach like a conventional graphical user interface. However, that would require information about the task domain, limiting a primary advantage of using a general purpose LLM.
To remain generalizable,  DirectGPT does not implement such custom interactions and relies only on general direct manipulation techniques.

\begin{figure*}[t]
\includegraphics[width=\textwidth]{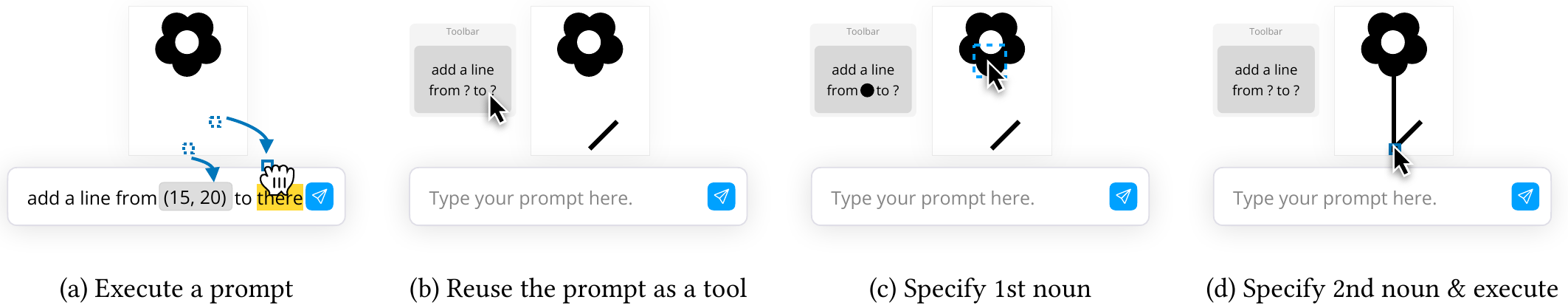}
    \caption{Prompts can be reused as tools: (a) a prompt with two nouns is executed; (b) the prompt is abstracted into an ad hoc tool; (c) using the tool, a click sets the first noun; (d) a second click sets the second noun and executes the reused prompt.}
    \label{fig:reuse}
    \Description{Screenshots of a scenario in which DirectGPT is used to to draw lines. First the line is drawn by prompting using locations drag and dropped into the prompt. Then the ad hoc line drawing tool is used by clicking twice.}
\end{figure*}

\subsubsection{Rapid Operations Through Prompts Reused as Tools}
Direct interfaces leverage labelled buttons to execute  commands~\cite{shneidermanDirectManipulationStep1983}.
It is difficult to predict what actions users may need, but a direct interface for LLMs can assist by creating an interface adapted to emergent tasks by reusing previous prompts.
Conveniently, composing prompts through physical actions makes explicit the verb and noun of an operations.
This means our system can reliably determine which parts of the prompt describe the objects being manipulated and which parts are concerned only with the action.
This allows abstracting prompts into universal commands, and an executed prompt can be reused immediately by preserving the verb, but swapping the noun.

In DirectGPT, as soon as a prompt finishes executing, it is added as a button in a toolbar. This prompt is now a \emph{tool}.
Pressing it enters a mode where further clicks on objects apply the prompt to these objects as well.
For example, when editing text, selecting a word and prompting ``synonym'' replaces the word with a synonym and adds a button labelled ``synonym'' to the toolbar (\cref{fig:synonym}b).
Then, this prompt can be reused as a tool.
First the user clicks on the tool, then selects one or multiple words.
As soon as the selection is complete, the prompt ``synonym'' is applied to them, behaving like tools in conventional direct interfaces (\cref{fig:synonym}c).

This technique scales to prompts involving multiple nouns.
For example, when editing an image, the user prompts ``add a line from here to there'' and then replaces ``here'' by dragging the first point, and ``there'' by dragging the second point (\cref{fig:reuse}a).
After drawing the line, the prompt is added in the toolbar as ``add a line from ? to ?'' (\cref{fig:reuse}b).
Clicking it selects the tool, so the next click on the canvas modifies the tool to show that point or object, such as ``add a line from (50, 50) to ?''. Then, a second click executes the prompt for the two specified locations (\cref{fig:reuse}c-d).

Tools also support noun-verb constructions without a mode.
The objects of interest can be selected before the tool is clicked.
In this situation, the tool is only used once and the mode is restored immediately after executing.
For example, a user can select two locations (by maintaining the ctrl key) and then click ``draw a line from ? to ?''. The line is drawn and the tool is no longer selected.

\subsubsection{Immediate Targeted Feedback}
Direct interfaces provide instant feedback so that \textit{``users can immediately see if their actions are furthering their goals, and if not, they can simply change the direction of their activity''}~\cite{shneidermanDirectManipulationStep1983}.
While LLMs will surely execute faster in the future, for now they may take seconds.
A direct interface can provide useful loading feedback so users can identify possible mistakes.
This is possible because direct manipulation disentangles nouns from verbs, so the interface is immediately aware of the objects that are to be edited.
This information can be used for targeted feedback about the specific objects that are being acted upon.

In DirectGPT, we use this information to show a loading animation as a pulse (a looping fade-in-fade-out) on objects selected before executing a prompt, or mentioned in the prompt.
While the feedback does not necessarily match the objects that will be modified (e.g., when referring to object as exception such as ``turn all circles blue except [this]''), it should match the user's mental model of their natural language query. Thus, this visual feedback lets the user verify the correct objects were selected and draws their attention to changes once they happen.
If the wrong objects are highlighted, the user can click a button to stop the generation.

\subsubsection{Reversible Operations Through Undo Mechanisms}
Direct interfaces make operations easily reversible to encourage exploration and reduce user anxiety~\cite{shneidermanFutureInteractiveSystems1982}.
The advantage is that the users' conceptual model of an operation might be clearer than in a conversational interface because it is similar to how typical software behave.

In DirectGPT, we implement a typical undo and redo mechanism.
Clicking the ``Undo'' button, or using the ctrl+z shortcut reverts the generated output to the state  before the last command.
Clicking the ``Redo'' button or using the shortcut restores the modification.
To align with the user's mental model, reverting matches the granularity of operations as performed by the user.
For example, if the user selects five elements and applies a prompt to them all at once, then this whole interaction is considered as one revertible operation.

\subsection{Implementation}\label{sec:directgpt_implementation}

\name works by first entering a prompt to generate an object of interest. Then, the object of interest is continuously represented and elements composing it can be manipulated as described in \cref{sec:directgpt}. Currently, \name supports all text-based representations such as text and code. It also supports SVG images by rendering them for direct selection of elements in the vector image.

DirectGPT is implemented in TypeScript using ReactJS~\cite{ReactJavaScriptLibrary2013} for the interface, Prism.js~\cite{PrismLibrary2015} for code syntax highlighting, and the official OpenAI API library~\cite{openaiOpenAINodeAPI2023} for executing prompts using  ``gpt-3.5-turbo''.
All source code and a live demo are available\footnote{Live demo and code: \url{http://ns.inria.fr/loki/DirectGPT}}.

This section describes the strategies we used to convert the direct manipulation actions described \cref{sec:physical_actions} into prompts.

\subsubsection{Localizing the Effect of a Prompt} The constraint of localizing the effect of a prompt is done by the interface. Instead of asking the LLM to rewrite the entire passage, the LLM is only asked to rewrite the part that is selected. Then, this part is replaced by the response from the LLM. This ensures a deterministic outcome. Note that the whole passage is still provided to the model with the selected part replaced by ``<blank>''. This ensures the LLM is aware of the context surrounding the selection. An example of this kind of engineered prompt is provided in \cref{appendix:localize}.

\subsubsection{Referring to Objects in Prompts} When a prompt contains objects that were dragged and dropped into the prompt field, our system  converts the prompt into one of two kinds of engineered prompts we empirically found to give the best results for text (including code, etc.) and vector images.

For text, simply copying the text that is being referred is not enough because the instruction would be ambiguous when the text appears multiple times. Instead, we use the delimiter strategy by copying the entire passage with added delimiters around the text that was selected by the user~\cite{openaiGPTBestPractices2023}. The delimiters all have a unique identifier, and this identifier is used in the engineered prompt in lieu of the object-words for text specified through direct manipulation. \cref{appendix:refer_text_obj} provides an example and more details.

For elements in a vector image, the delimiter strategy is not ideal because the code corresponding to an element is not always located within a contiguous part of the SVG specification. Instead, our system automatically adds unique ids to each SVG element, and the object-words representing shapes specified by direct manipulation are replaced by their corresponding id. \cref{appendix:refer_vector_obj} provides an example and more details. 

\newcommand{\baseline}{ChatGPT\xspace}

\section{User Study: DirectGPT vs ChatGPT}
We conducted a user study to measure the effect of direct manipulation principles to convey intent, control LLMs, recover from mistakes, and reuse prompts.
Participants edited literary text, JavaScript code, and vector graphics images using both \name and a replica of ChatGPT.
This replica of ChatGPT served as a baseline because it is one of the most popular interfaces to interact with LLMs (with estimations of more than 100 million ChatGPT users in 2023~\cite{milmoChatGPTReaches1002023}) and it is strictly conversational: a conversation is initiated with a prompt, answers are shown word-by-word, and the entire conversation is kept at all times and fed back into the LLM. 

For the purpose of the study and to ensure accurate logging, the ChatGPT interface was reimplemented.
Care was taken to ensure a faithful reimplementation: the prompting and conversation interface were made to be identical, the conversation panel rendered Markdown and supported code syntax highlighting, the system prompt was identical to the one used in ChatGPT, and the answers were streamed to display word-by-word.
We confirmed faithfulness by running all our experimental conditions using the official ChatGPT and comparing the responses.
Results were sensibly identical\footnote{LLMs are non-deterministic even at low temperatures, and some variations for identical prompts can be expected}.
The interface only differed in that it was not possible to start a new conversation or have multiple conversations in parallel, and our replica rendered SVGs mentioned in the conversation to avoid the need for an external SVG viewer.
This also meant that when an SVG was rendered, its code could not be seen by the participant.
This was done to make tasks involving images feasible and ensured participants manipulated images rather than code (which is tested separately).
For the sake of clarity, we refer to this interface as ChatGPT and use OpenAI's ChatGPT otherwise.

The \name condition used the same interface as the ChatGPT replica but with the direct manipulation principles described in \cref{sec:directgpt}.
Both conditions used the same model ``gpt-3.5-turbo'' accessed through the OpenAI Chat API~\cite{openaiOpenAIPlatform2023}.

\subsection{Participants and Apparatus}
We recruited 12 participants from our institution (20 to 34 age range, M=26.8, 6 self-identified as female and 6 as male).
Potential participants were asked if they had prior experience with programming and OpenAI's ChatGPT, they had to answer ``yes'' to both questions to participate. 
On a 5-point scale from 1-``never'' to 5-``often'' they reported their frequency of use of ChatGPT as Mdn=3 and having used AIs in the past to do text editing (N=6), code editing (N=8), and image generation or editing (N=8).

Participants took part in the study remotely. They shared their screen with the experimenter who took notes during the session. The participants' screen, the audio, and their interactions with the tool were recorded. Sessions were an hour long. In appreciation for their time, participants received the equivalent of \$11 USD.

\begin{figure*}[t]
    \includegraphics[width=\textwidth]{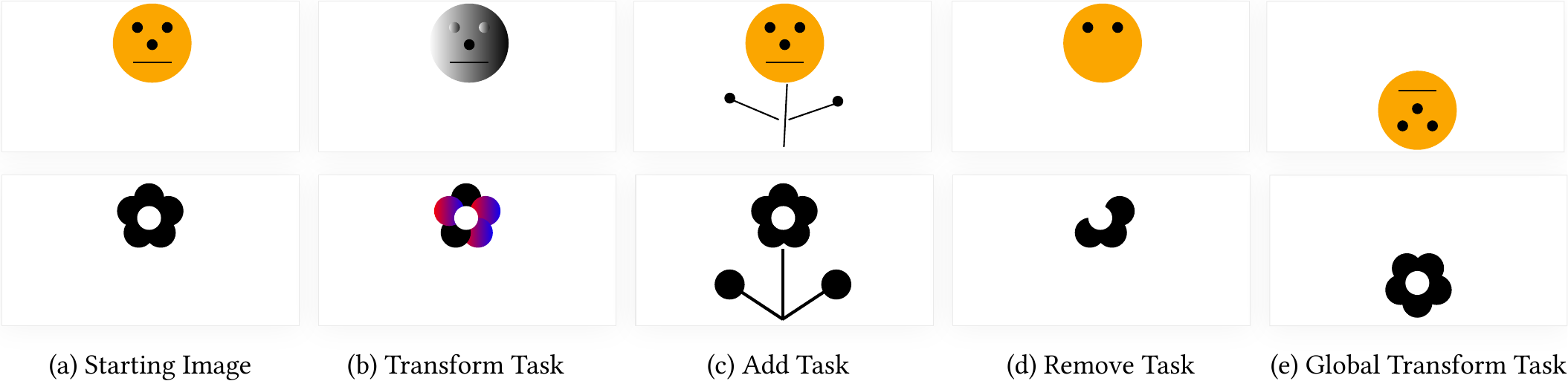}
    \caption{Given the (a) starting image, the participant completed four tasks in the image \textsc{activity} with ChatGPT and DirectGPT from either the top or bottom row: (b) colourize using gradients; (c) add elements; (d) remove elements; (e) flip upside down.}
    \label{fig:image_task}
    \Description{Images that the participant had to reproduce during the study. On the top row are tasks for the smiley. On the bottom row are tasks for the flower.}
\end{figure*}

\subsection{Procedure}

The participant used two \textsc{interface} (ChatGPT, \name). For each \textsc{interface}, they completed three \textsc{activity} (text, code, image). For each \textsc{activity}, they performed four tasks.\medskip

\noindent\textit{Introduction ($\sim$2min) --} After completing the consent form and the demographics questionnaire, the participant was told ``You will interact with an AI to edit text, code, and images''.\medskip

\noindent\textit{Video Tutorial (}2 \textsc{interface} $\times$ 3 \textsc{activity} $\times$ \textit{30sec) --}
Before each \textsc{activity} (including those done with the ChatGPT interface), the participant watched a video tutorial with a voice-over to describe the interactions possible with the interface they were about to use.
The video tutorials were designed to avoid giving away strategies or biasing participants in their wording of prompts.
This was done by showing toy examples (e.g., Lorem ipsum text) and demonstrating interactions with descriptive prompts (e.g., ``A prompt that modifies only the selected text''). These video tutorials are provided as supplementary material.\medskip

\noindent\textit{Editing Tasks (}2 \textsc{interface} $\times$ 3 \textsc{activity} $\times$ \textit{$\sim$8min) --}
After watching the video tutorial, the participant completed the four tasks of the \textsc{activity}.
The participant proceeded through the tasks one after the other. The current task to complete was displayed at all times on the left of the interface.
It consisted of a short textual instruction (e.g., ``Reproduce'' or ``Text in yellow => synonyms'') that could not be selected, and an image of the content to edit with some text or code in yellow, or the image the participant had to reproduce.
Because all tasks required editing existing content, this content was already added to the conversation as the first message to emulate an already first generation and make sure the task was identical across participants. This content was reset between tasks.
After completing a task, either because the participant finished, abandoned, or reached a time limit of three minutes, the participant responded to a 5-point semantic differential scale ``How close are you to the target'' from ``distant'' to ``close''. \medskip

\noindent\textit{Questionnaire on Interface Used (}2 \textsc{interface} $\times$ \textit{$\sim$2min) --} After completing all three \textsc{activity} for an \textsc{interface}, the participant completed a questionnaire including a System Usability Scale (SUS)~\cite{brookeSUSQuickDirty1995} and 5-point scale statements to rate. The questionnaire was identical after both \textsc{interface}. \medskip

\noindent\textit{Semi-structured Interview ($\sim$10min) --} At the end of the study, the participant was invited to comment on any aspect of the study. Then, the experimenter prompted the participant on their preferred interface, the most difficult tasks, prior experiences with AI where they faced similar difficulties, their strategies to write prompts, and other behaviours that the experimenter noticed during the session.

\subsection{Design and Tasks}

To test the flexibility of our approach, participants edited content through three activities: literary text, JavaScript code, and vector images.
The texts were the first three paragraphs from \textit{Alice's Adventures in Wonderland} by Lewis Carroll (253 words) and the first two paragraphs from \textit{Frankenstein; or, The Modern Prometheus} by Mary Shelley (265 words). 
The codes were a function to print a pyramid in the console (13 lines, a for loop containing two nested for loops) and a function to count the number of values below the mean of a moving window (15 lines, a for loop containing two nested for loops).
The images were a flower (5 black circles for the petals and 1 white circle in the centre, \cref{fig:image_task}) and a smiley face (1 yellow circle for the face, 3 black circles for the eyes and nose, and a black line for the mouth, \cref{fig:image_task})\footnote{Tasks and activities can be tested in the live demo: \url{http://ns.inria.fr/loki/DirectGPT}}.

For each content to edit, we designed four tasks that spanned different level of difficulties in using the direct manipulation features. Three tasks required localized edits (either modify specific elements, add/expand elements, or remove/reduce elements), and the fourth task required a global modification. We also ensured the tasks where within the boundaries of what the LLM could accomplish if prompted correctly. This was done to not confound our results with the limitations of the underlying model in accomplishing complex tasks.
For the text activity, the tasks were: replacing 5 words by synonyms; adding more descriptions to two passages; summarizing two passages; and using the future tense throughout the text excerpt.
For the code activity, the tasks were: renaming a variable; converting two for loops into while loops; factorizing a loop by using the reduce or repeat function; and converting the function to Python.
For the image activity, the tasks were: colouring three elements using a gradient; adding three lines and two circles (either to add a torso and hands to the smiley face, or to add a stem and leaves to the flower), removing three elements; or turning the image upside down (\cref{fig:image_task}). 

The study followed a within-subject design. \textsc{interface} (DirectGPT or ChatGPT) and \textsc{activity} (text, code, or image) appeared an equal number of times in each position.
This was done using a Graeco-Latin square to order \textsc{activity} and the version A or B of the content to edit.
Essentially, this means each participant had a unique ordering, half the participants started with ChatGPT, and the version A or B of the content to edit appeared equally as often with the ChatGPT interface than with the DirectGPT interface.

We gathered interaction logs, subjective ratings, and task measures for 2 \textsc{interface} $\times$ 3 \textsc{activity} $\times$ 4 task $\times$ 12 \textsc{participant} = 288 tasks.

\subsection{Data Analysis}
Considering ongoing debates related to statistics in HCI~\cite{dragicevicFairStatisticalCommunication2016, groupTransparentStatisticsGuidelines2019}, we report both p-values and 95\% confidence intervals on mean differences.
Because we are interested by the size of the effects rather than their mere existence, our interpretation is based on CIs that give information about the size of the
effects and the uncertainty around them. 
All 95\% confidence intervals are calculated using the studentized bootstrapping method because it has been shown to be the most robust across distributions and for study designs similar to ours~\cite{zhuAssessingComparingAccuracy2018, massonStatslatorInteractiveTranslation2023}. All p-values are calculated using a Wilcoxon signed-rank test because it is more robust and less likely to yield false positives compared to parametric alternatives when dealing with data from unknown or possibly heavily skewed distributions~\cite{bridgeIncreasingPhysiciansAwareness1999}.
Our analysis is done in Python using scipy 1.10.1~\cite{scipy} and arch 5.5.0~\cite{kevin_sheppard_2023_7975104} for the bootstrapped CIs. The Wilcoxon signed-rank tests discarded ties. The studentized bootstraps used 10,000 replications.

\subsection{Results}
\begin{figure*}[t]
    \includegraphics[width=\textwidth]{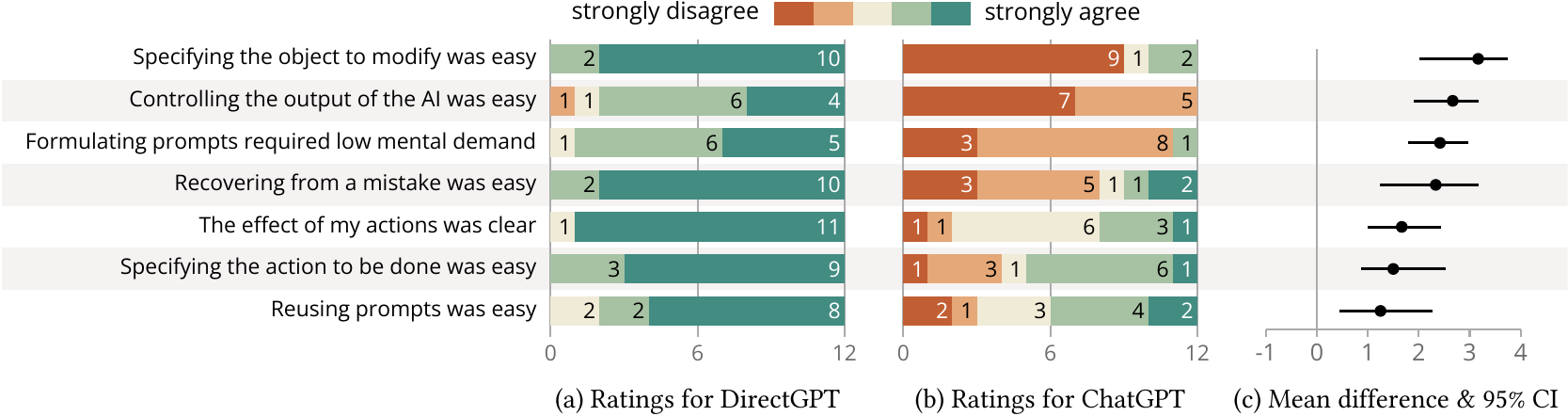}
    \caption{Participants' response when rating the 5-point statements for (a) \name and (b) ChatGPT. (c) Dots are the mean differences of DirectGPT compared to ChatGPT. Bars are the 95\% CIs calculated with the studentized bootstrap method.}
    \label{fig:statements}
    \Description{Stacked bar chart of participants responses to 5-point statements with the mean difference showing on the right. Overall, ratings for DirectGPT are mostly positive while those for ChatGPT are mixed or negative.}
\end{figure*}

\begin{figure*}[t]
    \includegraphics[width=\textwidth]{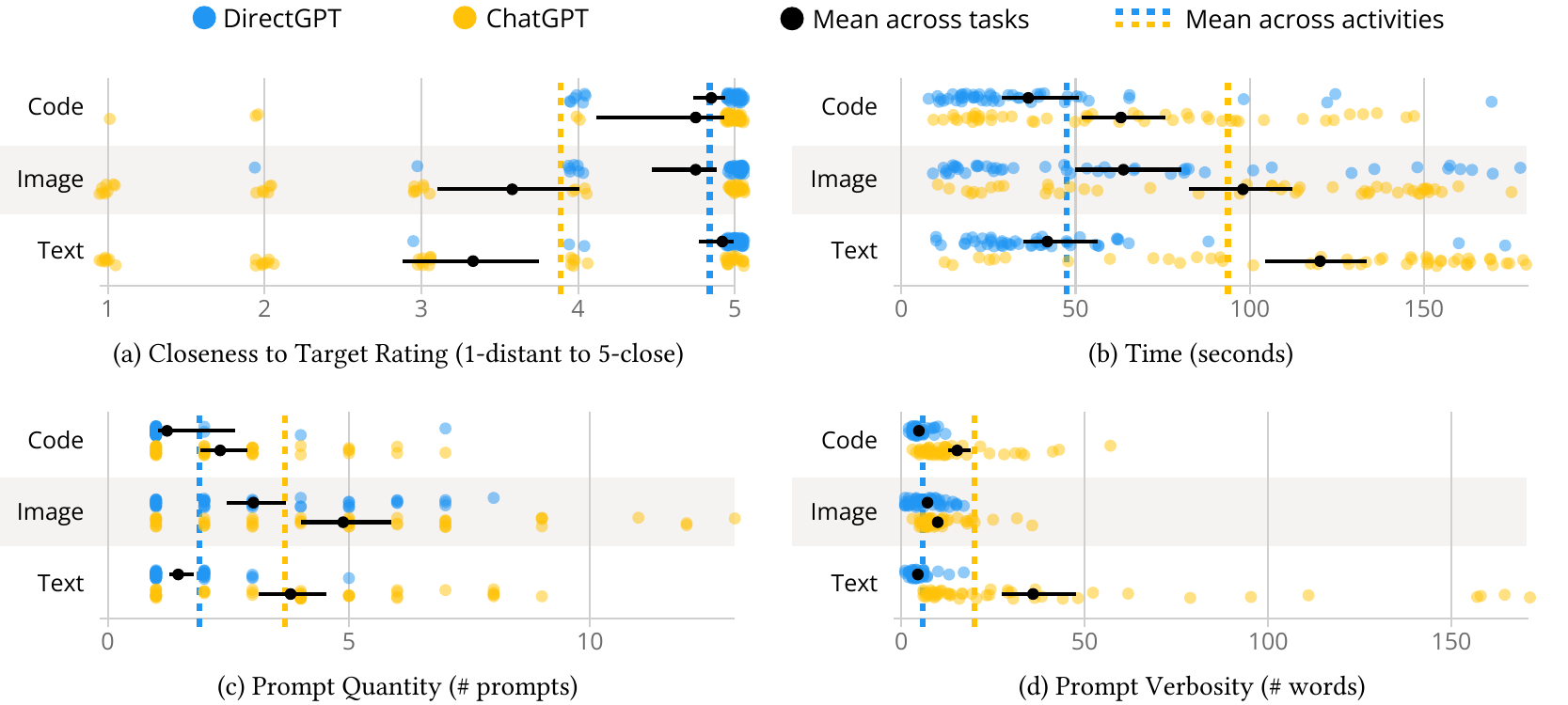}
    \caption{Result breakdown per \textsc{interface}, \textsc{activity}, and task for (a) closeness rating; (b) time; (c) number of prompts; (d) length of prompts. Black dots are means, bars are 95\% CIs, vertical dashed lines are means over all activities.}
    \label{fig:measures}
    \Description{Four scatter plots showing the distribution of the results. Overall, tasks in DirectGPT are better than in ChatGPT. Confidence intervals do not overlap except for Code in the closeness target rating and Code in the prompt quantity metric.}
\end{figure*}

Participants went through the tasks without abandoning. Averaged over all tasks and activities, participants consistently had better results with \name on objective metrics (time, number of prompts, number of words per prompt) and equal or better results on subjective metrics (closeness to target, usability, and 5-point scale statements). All results are detailed below and visualized in \cref{fig:statements} and  \cref{fig:measures}.


\subsubsection{Closeness to target} On a 5-point semantic differential scale, participants rated their modifications closer to the target when using \name by .95 {\small(95\% CI: [0.70, 1.25], M=4.84 vs M=3.89, p<0.001)}. Per \textsc{activity}, the mean difference for text was 1.58 {\small(95\% CI: [1.10, 1.99], p<0.001)} and for vector images it was 1.17 {\small(95\% CI: [0.71, 1.68], p=0.001)}. However, the mean difference for code was only 0.10 and likely insignificant {\small(95\% CI: [-0.13, 0.66], p=0.495)}.



\subsubsection{Time} The mean completion time of a trial was shorter with \name by \prettytimestamp{61.890} {\small(95\% CI: [\prettytimestamp{49}, \prettytimestamp{71.203}], M=\prettytimestamp{56} vs M=\prettytimestamp{117.709}, p<0.001)}. Even when excluding the time it took for the prompt to be executed and the output fully generated, it remains that \name was faster by \prettytimestamp{46} {\small(95\% CI: [\prettytimestamp{38}, \prettytimestamp{55}], M=\prettytimestamp{47} vs M=\prettytimestamp{93.537}, p<0.001)}. 
Per \textsc{activity}, the mean difference for text was \prettytimestamp{78} {\small(95\% CI: [\prettytimestamp{65}, \prettytimestamp{92}], p<0.001)}, 
for vector images it was \prettytimestamp{34} {\small(95\% CI: [\prettytimestamp{20}, \prettytimestamp{45}], p=0.001)}, 
and for code it was \prettytimestamp{27} {\small(95\% CI: [\prettytimestamp{3}, \prettytimestamp{39}], p=0.007)}.
Note that these include times from tasks where participants had to stop because they reached the 3min time limit. This happened in 8\% of the trials with \name, and 35\% with \baseline.

\subsubsection{Prompt Quantity} The mean number of prompts was lower with \name by 1.77 {\small(95\% CI: [1.37, 2.46], M=1.90 vs M=3.67, p<0.001)}. 
Per \textsc{activity}, the difference for text was of 2.33 {\small(95\% CI: [1.61, 3.40], p<0.001)},
for images it was of 1.88 {\small(95\% CI: [1.06, 3.26], p=0.002)},
and for code it was of 1.10 {\small(95\% CI: [0.56, 1.44], p=0.001)}.

\subsubsection{Prompt Verbosity} Prompts in \name were less verbose on average by 15.3 words {\small(95\% CI: [9.35 28.72], M=5.83 vs M=19.98, p<0.001)}.
Per \textsc{activity}, the difference for text was of 37.97 {\small(95\% CI: [21.59, 78.37], p<0.001)},
for code it was of 10.71 {\small(95\% CI: [6.13, 31.83], p<0.001)},
and for images it was of 3.21 {\small(95\% CI: [1.77, 4.89], p=0.002)}.

\subsubsection{Usability} On the System Usability Scale, and for the specific given tasks, \name was rated 92 (typically considered ``Excelent''~\cite{bangorEmpiricalEvaluationSystem2008}) whereas \baseline received a 53 (typically considered ``OK''). This corresponds to a mean difference of 39 {\small(95\% CI: [27.3, 47.34], p<0.001)}.

\subsubsection{Questionnaires} Participants rated all seven 5-point statements higher for \name. Ranked from largest to smallest effects: ``Specifying the object to modify was easy'' by 3.17 {\small(95\% CI: [2.01, 3.74], p<0.001)},
``Controlling the output of the AI was easy'' by 2.67 {\small(95\% CI: [1.91, 3.17], p<0.001)},
``Formulating prompts required low mental demand'' by 2.42 {\small(95\% CI: [1.79, 2.97], p<0.001)},
``Recovering from a mistake was easy'' by 2.33 {\small(95\% CI: [1.25, 3.17], p=0.005)},
``The effect of my actions was clear'' by 1.67 {\small(95\% CI: [0.99, 2.44], p=0.004)},
``Specifying the action to be done was easy'' by 1.5 {\small(95\% CI: [0.88, 2.53], p=0.004)},
``Reusing prompts was easy'' by 1.25 {\small(95\% CI: [0.44, 2.27], p=0.016)}.

\subsubsection{Interactions Used} 
With \name, participants relied on direct manipulation mechanisms to craft 84\% of their prompts. Specifically, 68\% of all prompts were localized, 14\% contained at least one reference to an object dropped through direct manipulation, and 20\% were the result of using a tool generated from previous prompts (i.e., reusing a previous prompt).

\section{Discussion}
After discussing our results, we open the discussion to themes stemming from participant comments and experimenter observations.

\medskip\noindent\textit{\name helps convey intent unambiguously.} This was reflected in participants being 50\% faster to accomplish tasks with 50\% fewer and 72\% shorter prompts. Subjectively, participants unanimously agreed that ``Specifying the object to modify was easy''. In fact, compared to \baseline, this is the statement with the largest positive mean difference.

\medskip\noindent\textit{\name helps control and notice the effect of prompts.} This increase in control is reflected in participants achieving results they judged closer to the target and rating that controlling the output of the AI was easier with \name than \baseline.  On average, participants also found the effect of their actions to be clearer with \name.

\medskip\noindent\textit{\name helps error recovery and might help reusing prompts.} Participants generally found that recovering from a mistake was easier with \name, due to the undo feature. They used the toolbar for 20\% of all prompts and generally rated reusing prompts to be easier with \name. However, this was also the smallest difference since half of the participants also found reusing prompts to be easy with \baseline.

\medskip\noindent\textit{\name preserves prompting-only capabilities and people sometimes prefer them.}
All participants completed global tasks without using direct selections. This confirms that \name preserves prompting possibilities of \baseline and that people may switch between prompting and direct manipulation depending on the task and perceived benefit.

\subsection{Observations and Participants' Comments}

\medskip\noindent\textit{Prompts were more concise (and less polite) with \name.} Corroborating existing findings~\cite{zamfirescu-pereiraWhyJohnnyCan2023, ohLeadYouHelp2018}, eight of our participants wrote prompts in a human-to-human conversation style when using \baseline. For example, they asked politely (P4, P5, P6, P7); thanked the model (P5, P4); encouraged it such as \emph{``yes''} (P0), \emph{``Great. Now also change [...]''} (P7), \emph{``Awesome. You forgot the hands though. [...]''} (P5); used caps lock, smileys, or exclamation marks to emphasize some points (P4, P5, P7); asked questions to make the model notice the problem (P0, P2, P4, P6); and generally used more verbose language such as \emph{``actually I was expecting [...]''} (P3), and \emph{``similarly, [...]''} (P8).

These participants did not use that kind of language when using \name despite doing similar tasks and using the same underlying model.
During the semi-structured interview, they explained that \name understood them faster and did not feel like a chat. 
\begin{quote}
    \textit{``[with DirectGPT] I knew that by highlighting the specific parts I can just, like, put in very concise prompts and it will know exactly what to do... so it did not feel necessary for me to type in a whole conversation'' --- P7}
\end{quote}

\begin{quote}
    \textit{``[with DirectGPT] it was more like a tool. Like I did not need to make it comprehend anything. I just wanted to tell it what to do [..] it did not feel like an assistant with whom I have to converse'' --- P5}
\end{quote}

\medskip\noindent\textit{With \baseline, participants carefully checked the text generation because they doubted its accuracy.} Besides rating that the effect of their actions was less clear with \baseline, participants also felt it was difficult to verify its responses. Specifically, participants commented about the difficulties to check that nothing else had been modified when \baseline responded with the modified text. 
\begin{quote}
    \textit{``[with ChatGPT] verifying that it actually only did the thing that you wanted it to do, it's a bunch of text... basically impossible'' --- P6}
\end{quote}
\begin{quote}
    \textit{``my reading speed is not that fast so... I couldn't really tell.. is it like correctly modified at the location I want to modify, and I am not very confident that it will generate the exact same thing from my experience'' --- P4}
\end{quote}

\medskip\noindent\textit{Code was easier to refer to than text and vector images.} 
While the difference in performance was large for text and images, participants performed similarly for tasks on code with both \baseline and \name, albeit they still required more time, more prompts, and more words with \baseline. When asked about the tasks they found most difficult with ChatGPT, most participants mentioned either text (N=8) or images (N=4). Some explained this choice due to how easy it is to be clear and specific with code, but how difficult it is with images and text due to the lack of structure. 
\begin{quote}
\emph{``The code is simpler to navigate. It makes it simpler to determine where it should go and tell the model where it should go. But if it's a big text it's like, ugh, talking about it in paragraphs and sentences and occurrences of words... horrible.'' --- P6}
\end{quote}
\begin{quote}
\emph{``in the code task... I think I talked about the first inner for loop. So you could specify based on ordering.'' --- P11}
\end{quote}
\begin{quote}
\emph{``I'd say the picture task was more challenging [...] I feel like with the code, I have enough coding knowledge that I can specify almost the exact location to do something'' --- P1}
\end{quote}

\medskip\noindent\textit{Participants knew and used a range of common and effective prompt engineering strategies.} Perhaps due to their prior experience with ChatGPT, we observed some of the prompt engineering strategies suggested in the OpenAI wiki~\cite{openaiGPTBestPractices2023}, such as splitting tasks into simpler subtasks (N=12), providing examples (P5), and specifying the desired output length (P9). 
Another effective strategy adopted by four participants was to agree on a common terminology to refer to elements prior to the task.
For example, in the image activity, P10 and P11 asked the model to give a label to the elements in the image and then referred to those labels; P8 gave an object an identifier and asked the model to do the same for other elements; and P4 asked for a description of the image and then reused the wording.

\medskip\noindent\textit{Participants were excited about \name and how it could help them.}
Overall, all participants mentioned a preference for \name. When asked about features they found most useful, localizing the effect was mentioned most frequently (P0, P1, P3, P4, P5, P6, P7, P9, P10), followed by the toolbar of reusable prompts (P2, P10, P11), dropping objects in the prompt (P2, P8), and undoing (P8). Some explained that they had previous bad experiences with OpenAI's ChatGPT and Github Copilot where selection features like in \name could have helped.
\begin{quote}
    \emph{``it was horrible to use [OpenAI's] ChatGPT for this kind of situation because you just had no control over the output. [...] Especially what is important is that the area of effect is just small, like I don't want it to change anything but that.''} --- P5
\end{quote}
\begin{quote}
    \emph{``in using Copilot I really would like to select'' --- P3}
\end{quote}
\begin{quote}
    \emph{``I use [OpenAI's] ChatGPT primarily to look over text I've written [...] often times it ends up changing the entire text which isn't really what I want, so, I think for this task it would be really nice to have the first system [DirectGPT]'' --- P8}
\end{quote}
\begin{quote}
    \emph{``There are tools that start to integrate stuff but they are very limiting or expensive or very specific and this one seems like it is generic in a way that it's like basically [OpenAI's] ChatGPT but you can target it better''} --- P6
\end{quote}
Other participants were surprised because they thought ChatGPT was already pretty good.
\begin{quote}
    \emph{``I was actually surprised by the second system [DirectGPT] cause I thought like, the first system [ChatGPT] was already pretty well integrated... But I guess the second system [DirectGPT] did surprise me and when I was using it I was like oh wow, this will actually make my experience with LLMs so much better''} --- P7
\end{quote}

\subsection{Limitation}

\noindent\textit{The results of the study might be LLM dependent.}
While both interfaces used the same underlying model, \baseline arguably suffered the most from model misunderstandings.
For example, with the image activity of the study, some participants tried prompting the model to ``remove the nose''  but the model would often misunderstand and also remove the eyes of the smiley face.
However, even assuming a model with perfect accuracy, previous Wizard of Oz experiments found that people still prefer a combination of direct manipulation and natural language, especially when some objects are hard to describe~\cite{hauptmannSpeechGesturesGraphic1989}.
This result is consistent with previous work showing the benefits of combining gestures with language~\cite{kobsa86-xtra, boltPutthatthereVoiceGesture1980} and that more expressive ways of communicating  reduce misunderstandings compared to just text~\cite{daftOrganizationalInformationRequirements1986, lengelSelectionCommunicationMedia1988}.

\medskip\noindent\textit{Participants' prior experiences probably helped them perform better.} All our participants had experience with LLMs. This was clear from their use of effective prompt engineering strategies, as mentioned previously. Additionally, our participants also had experience in coding and could generally be considered expert computer users.
It is not clear whether less technical users would perform equally well with \name.

\medskip\noindent\textit{Direct manipulation might not be as useful for other tasks.} For instance, some selections might be easier to specify in the prompt such as ``all red circles'' rather than selecting all circles manually. Additionally, our study focused specifically on editing tasks and the need for control over the output. However, control is not always necessary, and there are some benefits in having the AI lead the way~\cite{guzdialFriendCollaboratorStudent2019, clarkCreativeWritingMachine2018, lehmannSuggestionListsVs2022}. Similarly, it is possible that other tasks that are more exploratory would not benefit from a direct manipulation interface because users do not have a specific target in mind. In this case, \name remains usable the way one would use ChatGPT by typing prompts without any selection.

\medskip\noindent\textit{The design of DirectGPT might be misleading.} For instance, the pulsing feedback matches the elements mentioned in the prompt, which, for complex queries, might differ from the ones that will be modified once the command is executed. To resolve this discrepancy, an approach could be to decompose the prompt by first asking the LLM to list the elements that will be modified and highlighting those. Additionally, because \name shows only the final output, the modification is not explained. While this was not an issue with the tasks of our study, some applications may require presenting the explanation to the user, either on the side or on-demand, to help with transparency.

\subsection{Future Work}

\noindent\textit{Testing other benefits of direct manipulation such as learnability and explorability.} Shneiderman cites other benefits of direct manipulation interfaces such as enabling a multi-layered approach to learning~\cite{shneidermanDesigningUserInterface2017}. As such, \textit{``novices can learn basic functionality quickly''} and \textit{``experts can work extremely rapidly to carry out a wide range of tasks, even defining new functions and features''}~\cite{shneidermanDirectManipulationStep1983}. Future work could investigate if such learning occurs with \name. From our observations, we expect that localizing prompts and referring to objects would be quickly learned but that reusing prompts would be more of an expert feature. Additionally, \name follows the guidelines for explorable interfaces, such as making the effects of actions visible and making it safe to experiment~\cite{draperLearningExplorationAffordance1993, massonSuperchargingTrialandErrorLearning2022, demulLearningUserInterfaces1996}. Future work could investigate the impact of these features on users' exploratory behaviours.

\medskip\noindent\textit{Exploring extensions to direct manipulation such as demonstrational interfaces and instrumental interaction.} Direct manipulation has been extended in many ways that could also benefit \name. For example, instrumental interaction~\cite{beaudouin-lafonInstrumentalInteractionInteraction2000, beaudouin-lafonDesigningInteractionNot2004} defines \textit{interaction instruments} as mediators between users and objects of interest. An application of this idea would be to \textit{reify}~\cite{beaudouin-lafonReificationPolymorphismReuse2000} prompts into interaction instruments that can be applied to different objects and that are easily reused and modified.
Another extension of direct manipulation that seems promising for LLMs are demonstrational interfaces~\cite{myersDemonstrationalInterfacesStep1992}. Demonstrating the operation would allow not only to specify nouns but also verbs through physical actions. For example, users provide examples of the action they want to perform (e.g., they replace a word by a synonym) and the system would derive a tool from it. The examples could even be used in the prompt generated by the system~\cite{openaiGPTBestPractices2023}. 

\medskip\noindent\textit{Using prompts to help select objects.}
Object selection in \name is either done solely by direct manipulation or solely by describing the objects in the prompt. Future work could investigate the design of an hybrid approach where specific prompts such as ``all red circles'' would form a selection that can then be refined through direct manipulation. One design for this feature could be to offer an additional prompt input field for the purpose of object selection. Another approach is to implement \textit{hooks} that analyze the prompt while it is being typed and detect references to objects. For example, while typing ``move red circles to the top'', the part ``red circles'' would be detected and highlighted by the system. Then, the user could click on it to select the circles and possibly modify the selection through direct manipulation. Besides references to objects, these ``hooks'' could also detect other elements such as colours, fonts, styles and offer widgets to edit the prompt through direct manipulation, similar to code projections~\cite{gobertLorgnetteCreatingMalleable2023}.

\medskip\noindent\textit{Testing alternative designs.}
\rev{The design of \name was chosen to best exemplify the principles of direct manipulation so that their benefit could be evaluated. While the interactions we proposed favoured simplicity and consistency, other designs could capture direct manipulation principles equally well and offer a different trade-off}.
For example, an alternative design could rely on contextual menus or popups with a prompt field appearing close to the selection. This would reduce travel time between selection and prompt field and could support chaining different prompt fields~\cite{wuPromptChainerChainingLarge2022}. However, this may clutter the screen, conflict with the software preexisting interaction mechanisms, and make transitions between global and local edits difficult once the user starts typing the prompt. Similarly, instead of undo features, the edits could be marked for the user to accept or reject, like ``track changes'' in editing software. Finally, more complex tasks might require prompting the model with a chain-of-thought~\cite{weiChainofThoughtPromptingElicits2022} which would require tuning the internal prompts used by \name and extracting only the relevant part of the answer. Ultimately, these design decisions should be decided based on the targeted tasks and application domain.

\medskip\noindent\textit{Exploring the use of \name for other tasks such as exploration, data wrangling, and GUI building.} We demonstrated \name to edit text, code, and vector images but the principles proposed extend beyond these use cases. First, not all tasks require editing. Instead, two prompt fields could be supported, one to edit the content in-place, the other to ask for details. In this case, the direct manipulation would help tailor the explanation to a piece of code, a specific topic mentioned, or an element of the image.
Second, even with editing, more activities could be supported as long as the objects of interest are manipulable and continuously represented in their final form. 
For example, one could imagine editing a website by clicking a panel and prompting ``make this responsive by shrinking the content''; doing data wrangling by manipulating a data table such as selecting a column and then prompting ``turn into a row''; stylizing a chart by selecting the axes and prompting ``use Open Sans''; arranging a graphical user interface by prompting ``align this and that horizontally'' and dropping buttons instead of ``this'' and ``that''; or selecting a row in a table from a rendered \LaTeX\xspace document and prompting ``use grey background''.

\medskip\noindent\textit{Using \name to control other generative models.} 
\rev{While \name is built atop an LLM, our findings point to the potential of direct manipulation to help interact with other kind of prompt-driven generative models.}
For example, a music track could be directly manipulated using its audio spectrum. This would allow interactions such as prompting to ``reproduce this part here using a saxophone instead''. Additionally, a limitation of the implementation of \name is that it needs to map physical actions to textual prompts. This sometimes requires workarounds such as using identifiers. In contrast, models that accept multi-modal inputs could help create more robust implementations of \name.

\medskip\noindent\textit{Tradeoffs of integrating \name into exisiting software.} \name was designed for easy integration into traditional software. Often, this could be done as simply as adding a prompt field somewhere in the interface, even if the software already has selection and dragging mechanisms. For example, in a drawing software, dragging a shape within the canvas would move it, while dropping it in the prompt field would allow referring to it. If such an integration is done, future work could investigate tradeoffs between the use of prompts and more traditional software operations. For example, it might be preferable to do precise tasks through direct interactions and reserve prompting for irregular and more complex tasks. But, akin to how expert features such as software shortcuts are not necessarily learned~\cite{cockburnSupportingNoviceExpert2015}, people might not always adopt the most efficient approach.

\section{Conclusion}
We characterized how the principles of direct manipulation can be leveraged to design interfaces for LLMs that help express intent, control the generation of output, and recover from mistakes.
This was demonstrated with \name, an interface for LLMs that can localize the effect of prompts, refer to objects of interest in prompts, reuse commands, and provide undo operations. In a user study, participants were faster, more successful, and they preferred to use \name instead of ChatGPT to perform text, code, and image editing tasks. 
Beyond these use cases, this work informs how traditional software could seamlessly integrate LLMs and other generative models to support co-creation with artificial intelligence.



\begin{acks}
This work was made possible by 
NSERC Discovery Grant 2018-05187,  
Canada Foundation for Innovation Infrastructure Fund 33151 ``Facility for Fully Interactive Physio-digital Spaces'', and
the LAI Réapp.
\end{acks}

\bibliographystyle{ACM-Reference-Format}
\bibliography{zotero_do_not_modify, references}


\appendix
\section{Appendix}

\subsection{Prompts Generated From Physical Actions}
Below, we describe the prompts generated from physical actions and that are sent to \textit{gpt-3.5-turbo}.

\subsubsection{Localizing the Effect of a Prompt}\label{appendix:localize}
Below is the prompt generated from the interaction described in the use case scenario where ``a White Rabbit'' is selected followed by the prompt ``add description of its tail''. The text is the second paragraph from \textit{Alice's Adventures in Wonderland} by Lewis Carroll.

\begin{dialogue}
\noindent So she was considering in her own mind (as well as she could, for the hot day made her feel very sleepy and stupid), whether the pleasure of making a daisy-chain would be worth the trouble of getting up and picking the daisies, when suddenly <blank> with pink eyes ran close by her.

\medskip<blank>: a White Rabbit

INSTRUCTION: add description of its tail

Rewrite <blank>. Follow INSTRUCTION

<blank>:
\end{dialogue}

The result should be only the specific selected part being rewritten. Then, \name takes care of incorporating it back into the overall text.

\subsubsection{Referring to Textual Objects in a Prompt}\label{appendix:refer_text_obj}
Below is the prompt generated from an interaction where two words from the text ``hot'' and ``suddenly'' were dropped in a prompt such that it reads ``replace [hot] and [suddenly] with synonyms''

\begin{dialogue}
\noindent So she was considering in her own mind (as well as she could, for the 0]hot0] day made her feel very sleepy and stupid), whether the pleasure of making a daisy-chain would be worth the trouble of getting up and picking the daisies, when 1]suddenly1] a White Rabbit with pink eyes ran close by her.

\medskip replace text delimited by 0] and text delimited by 1] with synonyms

Keep rest of the text identical
\end{dialogue}

This should result in the whole text being re-written. Note that the choice of delimiter was done to make sure the model would perceive it as a separate token and would not merge it with the surrounding text. This was empirically found to work slightly better than using more standard delimiters such as XML tags for example.

\subsubsection{Referring to Vector Objects in a Prompt}\label{appendix:refer_vector_obj} Below is the prompt generated from an interaction where the user prompted ``draw a line between this and that'' and then dropped two circles instead of ``this'' and ``that''

\begin{dialogue}
\noindent <svg width="300" height="150">

    \hspace{0.5cm}<circle cx="133" cy="33" r="5" id="c0"></circle>
    
    \hspace{0.5cm}<circle cx="151" cy="20" r="5" id="c1"></circle>
    
</svg>

\medskip Return modified SVG code to draw a line between element with id "c0" and element with id "c1"
\end{dialogue}

This should output the SVG rewritten with the line added. Note \name makes sure all elements have a unique \textit{id}. If not, or if the id is not unique, the ids are modified.

\end{document}